\documentclass[twocolumn,showpacs,preprintnumbers,amsmath,amssymb]{revtex4}
\usepackage{graphicx}
\usepackage{dcolumn}
\usepackage{bm}
\usepackage{amsmath}

\begin{document}

\title{Theory of dissociative recombination of a linear triatomic ion with permanent electric dipole moment: Study of HCO$^{+}$}
\author{Nicolas Douguet, Viatcheslav Kokoouline}
\affiliation{Department of Physics, University of Central Florida, Orlando, Florida 32816;\\
Laboratoire Aim\'e Cotton, CNRS, Universit\'e Paris-Sud XI, Orsay, France}
\author{Chris H. Greene}
\affiliation{Department of Physics and JILA, University of Colorado, Boulder, Colorado 80309}

\begin{abstract} 
We present a theoretical description of dissociative recombination of triatomic molecular ions having large permanent dipole moments. The study has been partly motivated by a discrepancy between experimental and theoretical cross-sections for dissociative recombination of the HCO$^+$ ion. The HCO$^{+}$ ion has a considerable permanent dipole moment ($D\approx$4 debye), which has not been taken explicitly into account in previous theoretical studies. In the present study, we include explicitly the effect of the permanent electric dipole on the dynamics of the incident electron using the generalized quantum defect theory, and we present the resulting cross section obtained. This is the first application of generalized quantum defect theory to the dissociative recombination of molecular ions.
\end{abstract}

\pacs{34.80.Ht}

\maketitle

\section{Introduction}
Dissociative recombination (DR) of small molecular ions such as HCO$^+$ or H$_3^+$  plays an important role in the chemistry of interstellar clouds and is therefore one of the key elements in modeling the composition and temporal evolution of the clouds. In recent years, the theoretical description of the DR of H$_3^+$ in collisions with low energy electrons \cite{kokoouline01,kokoouline03a,kokoouline03b,santos07} has exhibited good overall agreement with experimental data. Following the success of the theoretical method developed for H$_3^+$, a similar treatment \cite{mikhailov06,douguet08b} was applied to describe DR in HCO$^+$, motivated by the larger goal to develop a complete set of tools to treat dissociative recombination of small polyatomic molecular ions. 

The description of DR in HCO$^+$ has proved to be far from simple. Although the recent theoretical studies \cite{mikhailov06,douguet08b} have demonstrated that the non-Born-Oppenheimer Renner-Teller coupling plays an important role in DR of HCO$^+$ (similar to the role of Jahn-Teller coupling in H$_3^+$ DR), the theoretical cross-section obtained still remains about a factor of 2-3 smaller than the lowest experimental result. Here, we point out that different experiments with HCO$^+$ give quite different results for cross-sections and/or thermally-averaged rate coefficients of the process (see, for example, Fig. 3 of Ref. \cite{douguet08b}). Although the experimental data differ from each other by up to a factor of 4, it seems that existing DR theory is still missing some important effect(s). One of the possible improvements would be to account for the relatively large permanent electric dipole moment of HCO$^+$.  The permanent dipole moment of HCO$^+$ has been estimated in several studies \cite{woods75,botschwina93,yamaguchi94} with values ranging from 3.3 to 4.0 debye (1 debye is 0.393430 atomic units). In this study we adopt the value $D=3.9$ debye from Ref.  \cite{botschwina93}. Therefore this study is devoted to (1) the development of a theoretical approach that accounts for the electron interaction with the electric dipole moment in addition to the usual Coulomb interaction between the ion and the electron and (2) an application of the approach to DR  of HCO$^+$. We will use atomic units (a.u.) in this study.

One of the main theoretical techniques used in our previous studies is multichannel quantum defect theory (MQDT) \cite{seaton66,greene85,jungen96,aymar96}. The standard version of MQDT \cite{seaton66,aymar96} was initially developed for purely Coulomb potentials with centrifugal terms, i.e. potentials with asymptotic behavior $V(r)=-\frac{1}{r}+\frac{l(l+1)}{2r^2}$, where $r$ is the radial electronic coordinates and $l$ is an integer. A first generalization to the problem of a combined Coulomb and an attractive or repulsive dipole potential was introduced by Bely {\cite{Bely66} in the context of electron scattering from He$^+$, where the ionic dipole moment produces electron escape channels characterized by radial $r^{-2}$ potentials having noninteger or even complex values of $l$. Refs. \cite{greene79,greene82} have generalized MQDT to account for the interaction between colliding partners behaving asymptotically as $\frac{\alpha}{r^p}$, with arbitrary $\alpha$ and positive $p=0,1,$ or $2$.  Subsequent studies \cite{watanabe80, mies84,mies00,bogao08, bogao98,burke98} have developed applications to a broader class of long-range potentials, e.g. with $p=3,4$ and $6$ as well as to more general numerically-specified potentials.  In particular, the generalized theory (GMQDT) is well-suited for the present problem, where we want to account for the dipole-electron interaction in addition to the regular Coulomb potential at large electronic distances.  The GMQDT itself is developed in detail in Refs. \cite{greene79,greene82}, and thus we only summarize here the main insights into the physics involved in main formulas of GMQDT, as well as describing the adaptations to the theory that are needed in order to describe DR of molecular ions with a permanent electric dipole moment. As is shown below, the new theoretical DR method to account for the dipole moment is very similar to the one developed earlier \cite{mikhailov06,douguet08b} with comparatively small changes in the formulas for the reaction matrix $\hat{K}$ and in the closed-channel elimination procedure. Therefore, a DR approach similar to the one developed in our earlier study Ref. \cite{douguet08b} is employed here. For this reason, in this communication we discuss only the main steps of the approach and the differences with the previous study \cite{douguet08b}.

Electron scattering from neutral polar molecules was treated in a frame transformation methodology by Clark \cite{clark79} and tested quantitatively by Clark and Siegel \cite{clark80}.  That theoretical formulation and the subsequent work of Fabrikant\cite{fabrikant83} bears similarities to the present one, in that the combined dipolar plus centrifugal potentials for the scattering electron are first diagonalized to find the scattering eigenchannels.

\section{Hamiltonian of the problem}

In order to model the DR process of HCO$^+$, we write explicitly the Hamiltonian of the ion-electron system as $H=H_{ion}+H_{el}$, where $H_{ion}$ is the ionic Hamiltonian and $H_{el}$ describes the electron and its interaction with the ion. Here, $H_{ion}$ is basically the vibrational Hamiltonian of the molecular ion expressed in Jacobian coordinates  $\mathcal{Q}$=$\{ R_{\rm{CO}},R_{G\rm{H}},\theta,\varphi \}$, where $R_{\rm{CO}}$ and $R_{G\rm{H}}$ represent respectively the distances C-O and {\it G}-H ($\textit{G}$ is the center of mass of the CO bond), $\theta$ is the angle between vectors $\vec{\mathrm{OC}}$ and $\vec{G\mathrm H}$, $\varphi$ is the angle of azimuthal orientation of the bending. Here, we consider $R_{G\rm{H}}$ as an adiabatic coordinate representing the dissociation path. The Hamiltonian  $H_{ion}$ is explicitly written in Eq. (1) of Ref. \cite{douguet08b}. In this approach, we neglected the rotational motion of the CO bond in space, but included relative rotation of H and CO. This approximation is justified by a large CO/H mass ratio. However, the approximation may lead to an underestimated DR cross-section if the rotational autoionizing resonances in HCO play an important role in DR. As a result of the approximation, the projection $m_{\varphi}$ of the angular momentum $\hat L$ on the CO axis is conserved. We solve the Schr\"odinger equation for the ionic Hamiltonian keeping the $R_{G\rm{H}}$ coordinate fixed: This determines vibrational wave functions $\Phi_{ m_{\varphi},l}(R_{G\rm{H}};R_{CO},\theta,\varphi)$ and corresponding adiabatic energies $U_{m_{\varphi},l}(R_{G\rm{H}})$ that depend parametrically on $R_{G\rm{H}}$. Index $l$ here numerates the ionic vibrational states for a given $m_\varphi$.

For the electronic part  $H_{el}$  of the total Hamiltonian $H_{el}$ it was shown \cite{larson05a} that only electronic states $s\sigma $, $p\pi ^{-}$, $p\sigma $, and $p\pi ^{+}$  have a significant contribution to the HCO$^+$ DR cross-section. We include them into the treatment and, in addition, we also include the $d\sigma $ states having a relatively small effect of DR. Therefore, the electronic Hamiltonian $H_{el}(\mathcal{Q})$ for a fixed value of $\mathcal{Q}$ and integrated over all electronic coordinates can be written as an infinite number (Rydberg series) of matrices $5\times 5$ (see Eq. (1) of Ref.\cite{mikhailov06}). In the matrix, the Renner-Teller couplings between $p\pi ^{-}$, $p\sigma $ and $p\pi ^{+}$ states are explicitly taken into account via the coupling coefficients $\gamma$ and  $\delta$. On the other hand, in our previous studies \cite{mikhailov06,douguet08b}, no coupling between  $s\sigma $, $p\sigma $ and  $d\sigma$ states was accounted for (except at short range in the {\it ab initio} calculation of the potential surfaces) because these states are not coupled by the Coulomb field at large distances $r$ in the basis of electronic states with a definite angular momentum. Now, if we include the electron-dipole interaction as well, it will introduce a coupling between states with different electronic angular momenta. The coupling element between the   $s\sigma $, $p\sigma $ and  $d\sigma$ states (integrated only over the electron angular coordinates) behaves as $1/r^2$ at $r\to\infty$. To account such a behavior in the framework of ordinary (Coulomb field only) MQDT, one would have to use energy-dependent non-diagonal matrix element of the quantum defect matrix. For this reason, we cannot use the specified basis of electronic states to represent the electronic Hamiltonian of the system. However, we adopt the logic of \cite{clark79,clark80,fabrikant83} and utilize a different electronic basis for which the non-diagonal coupling elements would vanish at long range (or at least decrease with $r$  faster than $1/r^2$). 

We stress the two approximations used here: (1) We consider the dipole moment for all possible geometries of HCO$^+$ ion to be constant and equal to the permanent dipole moment $D$ of HCO$^+$ at its equilibrium position (linear configuration, $\theta=0$). (2) As we already mentioned, $d\sigma$ states have a minimal influence on the DR process and, therefore, we can neglect any coupling with the states and concentrate our attention uniquely on the coupling between $p\sigma $ and  $s\sigma$ states. In fact, we have verified the validity of the second approximation by comparing results with and without inclusion of couplings between $p\sigma $ and  $d\sigma$ states, and we find that the inclusion of the coupling with the $d\sigma$ states has a negligible effect on the final cross-section. To enhance the clarity of our presentation, we do not include the coupling with the $d\sigma$ states in the discussion below. With the mentioned approximations, the $s\sigma$-$p\sigma$ part of the electronic Hamiltonian $H_{el}(\mathcal{Q},r)$ has the form
\begin{equation}
H_{el}(\mathcal{Q},r)=\left( 
\begin{array}{cc}
-\frac{1}{r} &\frac{<Y_0^0|D \cos\theta_{e}|Y_1^0>}{r^2}\\ 
\frac{<Y_1^0|D \cos\theta_{e}|Y_0^0>}{r^2} &\frac{-1}{r}+\frac{1}{r^2}
\end{array}
\right) \ .
\label{eq:Hint}
\end{equation}
The above matrix is coupled to the two $p\pi^\pm$ states by the Renner-Teller coupling between $p\pi$ and $p\sigma$ states. The matrix elements for the coupling are given and discussed in Refs. \cite{mikhailov06,douguet08b}. 

Under the assumption that $D$ is constant for any $\mathcal{Q}$, the Hamiltonian depends only on $r$. The numerator in the non-diagonal elements is easily evaluated and we find $\tilde{D}=\langle Y_0^0|D \cos\theta_e|Y_1^0\rangle=\frac{D}{\sqrt{3}}$ (here the angle $\theta_{e}$ is the azimuthal angle of the electron in the molecular coordinate system). When we diagonalize the $2\times 2$ Hamiltonian above, we find two new electronic states  $|\tilde s\sigma\rangle$ and $|\tilde p\sigma\rangle$, which are superpositions of  $|s\sigma\rangle $ and $|p\sigma\rangle$ with a projection of the angular momentum still equal to zero. The radial functions related to these new channel eigenstates are solutions of two Schr\"{o}dinger equations with effective potentials different than for  $|s\sigma\rangle $ and $|p\sigma\rangle$ states:

\begin{equation}
\label{pot}
V_{\pm}(r)=-\frac{1}{r}+\frac{1\pm\sqrt{1+4\tilde{D}^2}}{2r^2}\,.
\end{equation}
The potentials $V_{+}(r)$ and $V_{-}(r)$ are respectively related to the channels $|\tilde p\sigma\rangle$ and $|\tilde s\sigma\rangle$. The effective potential for $|\tilde p\sigma\rangle$ is more repulsive than the one for $|p\sigma\rangle$; the effective potential for $|\tilde s\sigma\rangle$ is more attractive than the one for $|s\sigma\rangle$. We use the notations $|\tilde p\sigma\rangle$ and $|\tilde s\sigma\rangle$ for the new channel states because the mixing between $p\sigma$ and $s\sigma$ states is not very big: the new states are still mainly composed from their respective original states (around 75$\%$).

Changing electronic states from $|p\sigma\rangle$, $|s\sigma\rangle$ to $|\tilde p\sigma\rangle$ and $|\tilde s\sigma\rangle$, the $5\times 5$ electronic Hamiltonian $H=H_{el}(\mathcal{Q})$ including the five states $|\tilde p\sigma\rangle$, $|\tilde s\sigma\rangle$, $|d\sigma\rangle$, and $|p\pi^\pm\rangle$  has the same form as the one given in Refs. \cite{mikhailov06,douguet08b} if we change quantity $E_{\sigma}$ on Refs. \cite{mikhailov06,douguet08b} to $E_{\Tilde{\sigma}}$ to keep consistency in notations. The coefficients $\gamma$ and  $\delta$ are obtained from the {\it ab initio} potential energy surfaces $V_{\pi',\pi'',\sigma}(\mathcal{Q})$  \cite{larson05a} using the same formulas (3), (5), and (6) in Ref. \cite{mikhailov06}, where the states denoted as $p\pi'$ and $p\pi''$, are respectively symmetric and antisymmetric superposition of $|p\pi^{+}\rangle$ and $|p\pi^{-}\rangle$ with respect to reflexion in a plane containing the molecular axis. Therefore, the diagonalization of $H=H_{el}(\mathcal{Q})$ is accomplished by the same unitary matrix $\hat{U}$ as in  Eq. (2) of Ref. \cite{mikhailov06}. 

\section{Quantum defects and reaction matrix for the $-1/r\pm A/r^2$ potential}

One more difference with the treatment of Refs. \cite{mikhailov06,douguet08b}  is the way how the quantum defects functions $\mu_i(\cal Q)$ and the reaction matrix  $\hat{K}(\cal Q)$ are evaluated from the {\it ab initio} potential surfaces of HCO. The evaluation of quantum defects $\mu_i(\cal Q)$ in Refs. \cite{larson05a,mikhailov06,douguet08b} is made based on the Rydberg formula, which assumes an integer and non-negative partial wave quantum number and the asymptotic effective potential for the electron-ion interaction of the form  $V(r)=-\frac{1}{r}+\frac{\lambda(\lambda+1)}{2r^2}$ with $\lambda$ an integer usually denoted $l$. From Eq. (\ref{pot}) it is clear that the corresponding values of $\lambda$ are not integer numbers: they are obtained from the equation $\lambda(\lambda+1)=1\pm\sqrt{1+4\tilde{D}^2}$. Introducing the positive  $\Delta_{+}=5+4\sqrt{1+4\tilde{D}^2}$ and negative $\Delta_{-}=5-4\sqrt{1+4\tilde{D}^2}$ quantities, the new ``partial wave'' quantum numbers are
\begin{eqnarray}
\begin{array}{l}
\lambda_{\tilde p\sigma}=\frac{-1+\sqrt{\Delta_{+}}}{2}\,,\\
\lambda_{\tilde s\sigma}=\frac{-1\pm i\sqrt{-\Delta_{-}}}{2}\,,\\
\end{array}
\end{eqnarray}
$\lambda_{\tilde p\sigma}$ is real,  $\lambda_{\tilde s\sigma}$ is complex. Therefore, the standard quantum defect treatment cannot be applied. Below, we summarize the main ingredients of the generalized quantum defect theory as formulated in \cite{greene79,greene82}.

We start with an one-channel problem represented by the following equation, where $\lambda$ is a complex number:
\begin{equation}
\label{eq:channel}
\left(-\frac{1}{2}\frac{d^2}{dr^2}+\frac{\lambda(\lambda+1)}{2r^2}-\frac{1}{r}-\epsilon\right)f(\epsilon,\lambda,r)=0\,.
\end{equation}
The equation has two independent solutions behaving asymptotically for $\epsilon <0$ as $f^{\pm}(\epsilon,\lambda,r)\rightarrow e^{\mp \frac{r}{\nu}}r^{\pm \nu}$, where $\nu=\frac{i}{\kappa}=(-2\epsilon)^{-1/2}$ (For $f^\pm$ we use definitions of Ref. \cite{greene82}. See Eq. (3.7) of \cite{greene82}. In contrast, $f^\pm$ in Ref. \cite{aymar96} are defined differently, Eq. (2.38) in \cite{aymar96}). From these two functions the regular and irregular solutions for this equation (respectively denoted $f (\epsilon,\lambda,r)$ and $g(\epsilon,\lambda,r)$) can be constructed. In our case, the only basic requirement for $g (\epsilon,\lambda,r)$ is that it should lag $90^{\circ}$ with respect to the regular solution $f (\epsilon,\lambda,r)$. More generally, great care has to be taken to construct the irregular solution for certain types of potentials (as pure dipole or repulsive Coulomb potential) in a way that the solution remains independent on threshold effects. This construction procedure is discussed for instance in Ref. \cite{greene82}. The asymptotic expansion of these solutions as a superposition of $f^{\pm}$ is obtained using the following general formula. For instance, for $f (\epsilon,\lambda,r)$ we have:
\begin{eqnarray}
\label{superposition}
2ikf(\epsilon,\lambda,r)=\nonumber\\
W(f^{-},f)f^{+}(\epsilon,\lambda,r)-W(f^{+},f)f^{-}(\epsilon,\lambda,r)\,,
\end{eqnarray}
where the Wronskian $W(a,b)$ above is defined as $W(a,b)=a\frac{db}{dr}-b\frac{da}{dr}$. For the potential in Eq.(\ref{eq:channel}), the solutions are confluent hypergeometric functions. Using their asymptotic behavior at large $r$, we can write for $f$ \cite{greene79}:
\begin{equation}
\label{eq:f}
f (\epsilon,\lambda,r)=
(\pi\kappa)^{-\frac{1}{2}}(\sin\beta (D_\epsilon)^{-1} f^{-}-\cos\beta D_\epsilon f^+).
\end{equation}
In the above expression, the factor $D_\epsilon$  is a monotonic function of $\epsilon$ that rescales the amplitudes of $f^{\pm}$, and the oscillating factors $\sin\beta$ and $\cos\beta$ are parameters responsible for the energy-dependent mixing between the exponentially growing and decaying functions $f^{\pm}$. The wave function for a bound state should decay exponentially at infinity. Therefore, the condition for a bound state is simply $\sin\beta$=0 or equivalently $\beta$=$\pi(n+1$) with $n=0,1.2...$. If $\lambda$ is a real number,  $\beta$ is simply $\pi(\nu-\lambda)$. It corresponds to a pure Coulomb field with a centrifugal term and we recover the Rydberg formula for the energy of bound states $E=-0.5/(n+l+1)^2$. In a general case when $\lambda=\lambda_R+i\alpha$ is a complex number, the formula for $\beta$ is more cumbersome  \cite{greene82}
\begin{eqnarray}
\label{eq:beta}
\beta(\kappa,\lambda)=\nonumber\\
\pi(1/\kappa-\lambda_R)+\tan^{-1}[\tanh(\pi\alpha)\tan(y-\alpha \ln(2\kappa)]\,,\\
\mathrm{where}\  y=\arg(\Gamma(2\lambda+2)[\Gamma(\nu-\lambda)/\Gamma(\lambda+1+\nu)]^{1/2})\nonumber\,.
\end{eqnarray}
Having defined  the regular solution, we can write the irregular solution using a similar procedure. The irregular solution $g(\epsilon,\lambda,r)$ is then given by
\begin{equation}
\label{eq:g}
g (\epsilon,\lambda,r)=-(\pi\kappa)^{-\frac{1}{2}}(\cos\beta (D_\epsilon)^{-1}f^{-}+\sin\beta D_\epsilon f^+)\,.
\end{equation}

If we consider an additional short range potential in Eq. (\ref{eq:channel}), the asymptotic solution will in general be written as a superposition of regular and irregular solutions $F(\epsilon,\lambda,r)=f (\epsilon,\lambda,r)-Kg(\epsilon,\lambda,r)$, where $K$ is a mixing coefficient. From Eqs. (\ref{eq:f}) and (\ref{eq:g}) the asymptotic behavior of $F(\epsilon,\lambda,r)$ is easily derived. The requirement that the coefficient in front of the growing exponential must vanish in order for the solution to be physically acceptable determines $K$ as $K=-\tan(\beta)$.

Because HCO$^+$ has a dipole moment, evaluation of the reaction matrix $\hat K$ from {\it ab initio} energies has to be changed (in comparison with our previous study \cite{mikhailov06,douguet08b}). Namely, the reaction matrix $\hat{K}$ in the diagonal form and at fixed nuclear positions is written in terms of $\beta(\kappa,\lambda)$ as $K_{ii}=-\tan({\beta_i}(\kappa_i,\lambda_i))$ for each channel $i$ (we have five of them as discussed above). Index $i$ at $\beta_i$ means that we use different formulas depending on the asymptotic behavior of the potential in the corresponding channel. The value of $\kappa$ is obtained from the $\textit{ab initio}$ electronic energies of HCO and HCO$^+$ \cite{larson05a} with $n=4$ for $\tilde s\sigma$ state and $n=3$ for other states. The $\hat{K}(\mathcal{Q})$ matrix in the basis of states $\tilde s\sigma$, $p\pi ^{-}$, $\tilde{p}\sigma$, $p\pi ^{+}$ and $d\sigma$ is obtained using the unitary transformation as discussed in Ref. \cite{mikhailov06}: $\hat{K}(\mathcal{Q})=U^\dagger\tan(-\hat{\beta})U$.

\section{Cross-section calculation}

Similarly to our previous study, we solve the vibrational Schr\"odinger equation for HCO$^+$ fixing the adiabatic coordinate $R_{G\rm{H}}$ and obtain the eigenenergies $U_{m_{\varphi},l}(R_{G\rm{H}})$ (we call them adiabatic potentials) and the corresponding eigenstates  $\Phi_{ m_{\varphi},l}(R_{G\rm{H}};R_{\rm{CO}},\theta,\varphi)$ (adiabatic states). Then, we use $\Phi_{ m_{\varphi},l}(R_{G\rm{H}};R_{\rm{CO}},\theta,\varphi)$ to obtain the reaction matrix $\mathcal{K}_{\{m_\varphi,l,i\},\{m_\varphi',l',i'\}}(R_{G\rm{H}})$
\begin{equation}
\mathcal{K}_{\{m_\varphi,l,i\},\{m_\varphi',l',i'\}}(R_{G\rm{H}})=\langle \Phi _{m_\varphi,l}|K_{i,i'}(\mathcal{Q})|\Phi _{m' _\varphi,l'}\rangle\,,
\end{equation}
where the integral in the above expression is taken over the three coordinates, $R_{\rm{CO}}$, $\varphi$, and $\theta$.
The resulting reaction matrix $\mathcal{K}_{j,j'}$ ($j\equiv\{m_\varphi,l,i\}$) is multichannel and depends parametrically on $R_{G\rm{H}}$. It is then used to obtain for each $R_{G\rm{H}}$ value positions $U_a(R_{G\rm{H}})$ and widths $\Gamma_a(R_{G\rm{H}})$ of autoionizing resonances of the neutral HCO molecule. The procedure of resonance search is standard and is discussed, for example, in our earlier study \cite{kokoouline01}. The only difference is how the quantity $\beta(\kappa,\lambda)$ is calculated for states with non-integer $\lambda$.  $\beta(\kappa,\lambda)$ is needed to perform the closed-channel elimination procedure and evaluation of closed-channel mixing coefficients. (For more details about the channel elimination and the closed-channel mixing coefficients see Eqs. (2.50) and (2.54b) and the corresponding discussion in Ref. \cite{aymar96}). The quantity $\beta(\kappa,\lambda)$ should be calculated using Eq. (\ref{eq:beta}) for channels with non-integer $\lambda$. The obtained widths and positions of autoionizing resonances are used to calculate the electron-ion recombination cross-section as described in Ref. \cite{mikhailov06}. 
\begin{figure}
\includegraphics[width=8cm]{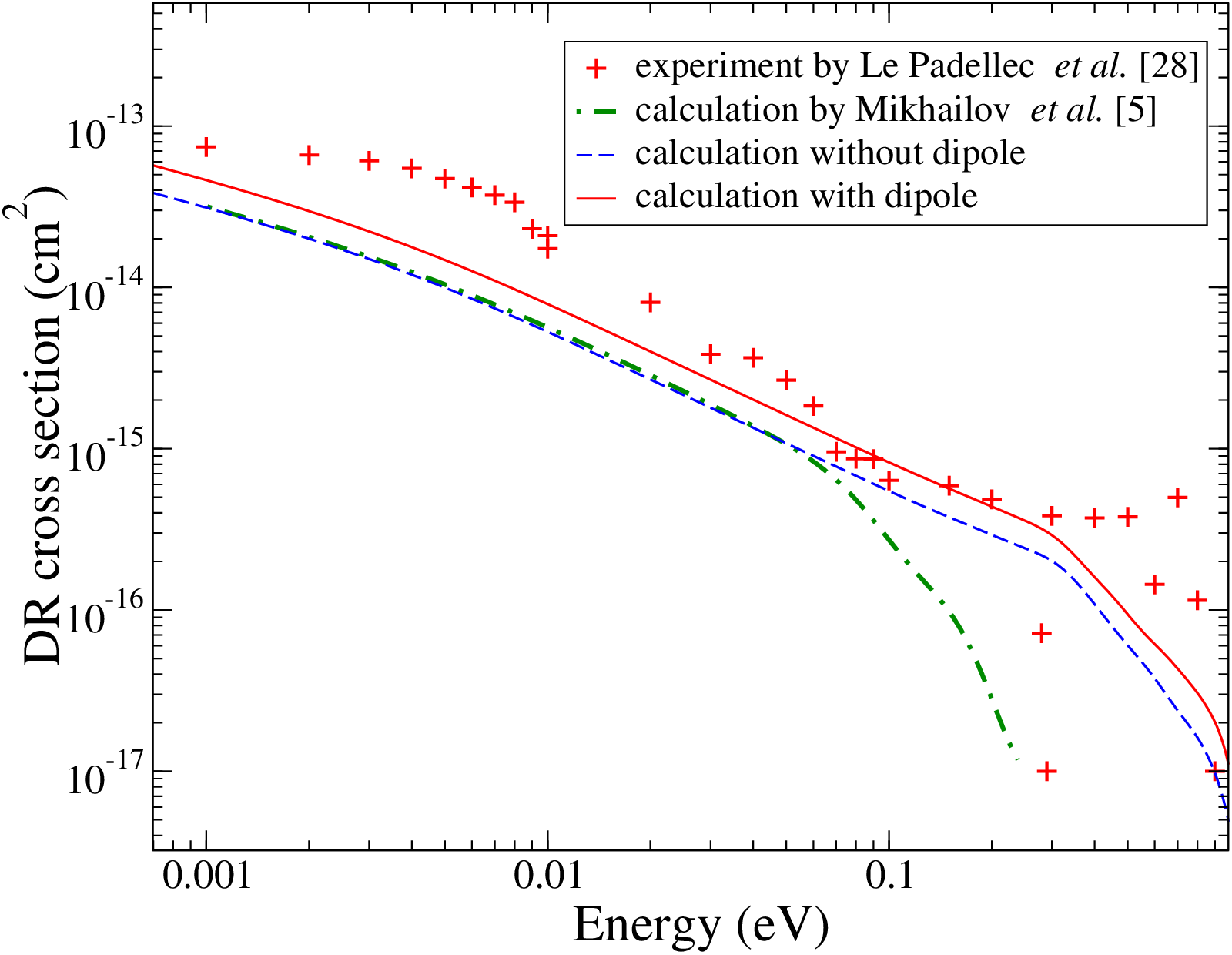}
\caption{(Color online) Calculated DR cross section for HCO$^+$ including dipole contribution (solid line) and the DR cross section for HCO$^+$ without dipole contribution as a function of the incident electron energy. The experimental \cite{lepadellec97} (cross symbols) and previous theoretical \cite{mikhailov06} (dashed line) cross sections are also shown for comparison. The theoretical curves include a convolution over the experimental electron energy distribution.}
\label{fig:merged-beam}
\end{figure} 

\section{Discussion of results}

Figure \ref{fig:merged-beam} shows the cross section obtained in the present study. The inclusion of the electric dipole moment of HCO$^+$ leads to an overall increase of about 50$\%$ for the cross section and agrees better with experiment than the cross section calculated neglecting the dipole moment. The agreement between theory and experiment is very good for electron energies between 0.06 eV and 0.3 eV. For larger energies, the experiment seems to have a large error bar (the experimental error bar is large possibly because the measured cross-section is small at higher energies). At energies below 0.06 eV the theoretical cross-section is below the experimental one by a factor of two approximately. The reason for the difference is not clear. One possible reason for the remaining disagreement could be related to the isomer HOC$^+$ of the HCO$^+$ if (1) the isomer is present in the experiment of Ref. \cite{lepadellec97} in a non-negligible fraction and (2) the DR cross-section for the isomer is significantly larger than the cross section for HCO$^+$. The preliminary calculation \cite{isomer-HOC} suggests that the ground ionic potential of HOC$^+$ crosses the dissociative potential of HOC in the Franck-Condon region. It means that the DR cross section for HOC$^+$ should indeed be larger and could be of order of a few 10$^{-7}$cm$^3/$s at 0.025 eV ($\sim$300 K). Nevertheless, this hypothesis must be verified in experiment and calculation.

We would like to discuss in some detail the increase of the cross section comparing with the results of the previous theoretical approach \cite{mikhailov06, douguet08b}. With the effect of the HCO$^+$ electric dipole moment included, it seems reasonable to expect a higher cross section due to an additional coupling between $s\sigma$ and $p\sigma$ states introduced by the dipole term. However, a more detailed analysis shows that the coupling is not directly responsible for the increase of the cross section, but it acts rather in an indirect way through the Renner-Teller effect as described below. The cross section in the present study is calculated using autoionizing resonances. Therefore, the cross section is increased if the density or/and widths of resonances become larger with the inclusion of the electric dipole moment. The density of resonances is expressed as $\frac{1}{\pi}\frac{d\beta}{dE}$. Therefore, we can compare the density of resonances by evaluating this quantity using the updated formula for $\beta$. For the $p$-states, the quantities $\beta$ for the pure Coulomb and Coulomb+dipole interactions differ from each other by an energy-independent term. Therefore, the density of resonances produced by closed $p$-wave channels is the same in the both treatments. For the $s$-states, analysis of the derivative of new $\beta(E)$  of Eq. (\ref{eq:beta}) gives only negligible differences for pure Coulomb and Coulomb+dipole interactions. Thus, the higher observed cross section is caused uniquely by an increase in widths of autoionizing resonances. By inspecting different resonances, we have indeed found that the increase in the cross section is related to larger widths of $p$-resonances. To apprehend this result, we consider non-diagonal elements of the $\hat{K}$ matrix. These terms can be written in the following way \cite{guisti80}:

\begin{eqnarray}
\label{eq:Ki}
K_{ii^{'}} (\mathcal{Q},E)=-\pi <\phi_{iE}\mid V \mid \phi_{i^{'}E}> +\nonumber \\
\Sigma_{k} {\cal  P}  \int dE^{'}\frac{<\phi_{iE} \mid V \mid \phi_{kE^{'}}>}{E-E^{'}}K_{ki^{'}}(\mathcal{Q},E')\,,
\end{eqnarray}
where indexes $i$, $i'$, and $k$ specify channels with the electronic wave function $\phi$, the symbol $\cal  P$ refers to the principal value of the integral, and $V$ is the potential of the system. Now in a first approximation, we can just retain the first term on the right of Eq. (\ref{eq:Ki}). As the width  of a Feshbach resonance is 
\begin{equation}
\Gamma =2\pi \frac{dE}{dn}\mid<\phi_{iE}\mid V \mid \phi_{i^{'}E}>\mid^2\,,
\end{equation}
where $n$ is the principal quantum number, we obtain
\begin{equation}
\label{eq:width}  
\Gamma\thickapprox\frac{2}{\pi n^3}\mid K_{ii^{'}} (\mathcal{Q},E)\mid^2\,.
\end{equation}
\begin{figure}
\includegraphics[width=8cm]{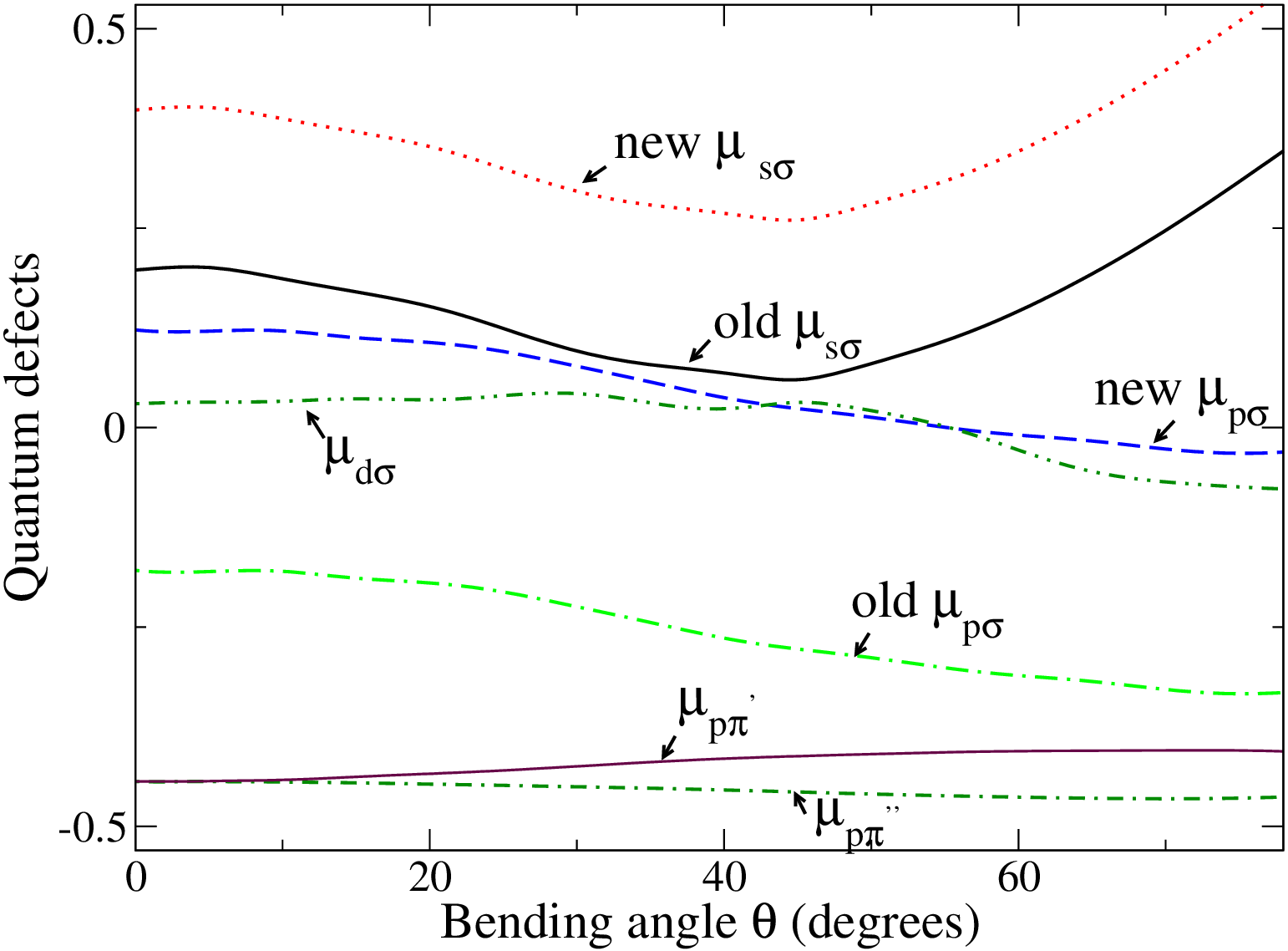}
\caption{(Color online) The figure shows quantum defects for the new states $\tilde{s}\sigma$ and  $\tilde{p}\sigma$ calculated from the {\it ab initio} electronic energies using $\beta$ of Eq. (\ref{eq:beta}) as well as the old quantum defect (calculated using $\beta$ determined by a pure Coulomb potential) as a function of the Jacobi angle $\theta$.  The two other Jacobi coordinates $R_{\mathrm{CO}}=2.00$ a.u. and $R_{G\mathrm{H}}=3.27$ a.u are fixed at their equilibrium positions for $\theta=0$.}
\label{fig:quantum defect}
\end{figure}
Because the widths are related to the non-diagonal elements of the reaction matrix $K_{ii^{'}} (\mathcal{Q},E)$, we can investigate the non-diagonal elements to understand the reason for the increase of the cross section. In the diagonalized form, the elements of the reaction matrix are $K_{jj'}=\tan (\pi\mu_j)\delta_{jj'}$ where the quantum defects $\mu_j$ are obtained directly from {\it ab initio} calculation for every internuclear configuration $\mathcal{Q}$ using  Eq. ( \ref{eq:beta}). The old quantum defect are also obtained from the same {\it ab initio} data but using the Rydberg formula. The new and old quantum defect are shown for comparison in Fig. \ref{fig:quantum defect} as a  function of the Jacobi angle $\theta$.
As it can be seen from the figure, the new quantum defect for the  $\tilde{s}\sigma$ and  $\tilde{p}\sigma$ states differ from each other by a simple translation. The translation, which has no direct effect on widths of the $\tilde{s}\sigma$ resonances (they are basically just shifted), has a drastic effect on the $\tilde{p}\sigma$ states coupled  to the $p\pi$ states by the Renner-Teller effect. It is possible to show that changing the basis for the reaction matrix from the eigenbasis (diagonal form of $K(\cal Q)$) to the basis of the five states $|\tilde p\sigma\rangle$, $|\tilde s\sigma\rangle$, $|d\sigma\rangle$, and $|p\pi^\pm\rangle$ discussed above, introduces a non-diagonal term $K_{ii'}$ between $\tilde{p}\sigma$ and $p\pi^{\pm}$ that is directly proportional to the difference $\tan(\mu_{\tilde{p}\sigma})-\tan(\mu_{p\pi'})$. From Fig. \ref{fig:quantum defect}, it is evident that the difference is increased for the new $\tilde{p}\sigma$ state comparing with the difference for the old (without dipole contribution) $p\sigma$ state. The coupling element between $p\pi^+$ and $p\pi^-$ are only slightly affected by the translation of the quantum defect.

\begin{figure}
\includegraphics[width=8cm]{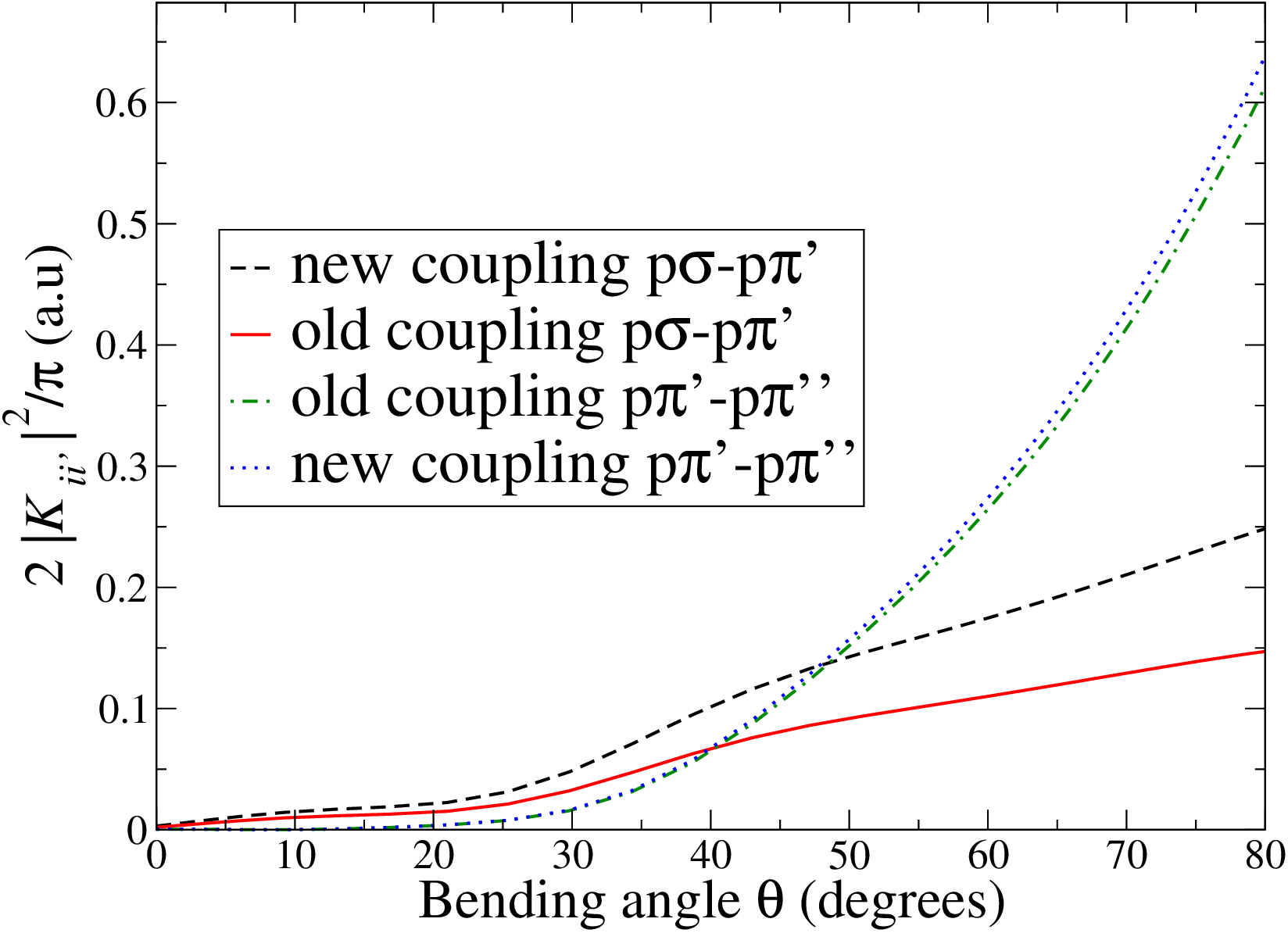}
\caption{(Color online) The figure shows the absolute value square of the non-diagonal elements $|K_{ii^{'}} (\mathcal{Q},E)|^2$ of the reaction matrix as a function of the bending angle $\theta$ around equilibrium values of $R_{\mathrm{CO}}$(2.00 a.u) and $R_{G\mathrm{H}}$ (3.27 a.u).}
\label{fig:width}
\end{figure} 

 The approximate width for the two coupling elements in the new and old can be estimated using Eq. (\ref{eq:width}). The result is shown in Fig. \ref{fig:width}. The coupling between $\tilde{p}\sigma$ and $p\pi^{\pm}$ is clearly increased compared to the previous treatment by about 50$\%$. On the other hand, the coupling  between the $p\pi^+$ and $p\pi^-$ states is changed only a little (less than 5$\%$). Thus, we conclude that the increase of the coupling between $\tilde{p}\sigma$ and $p\pi^{\pm}$ states is responsible for the higher cross-section in the present approach including the electric dipole moment of HCO$^+$. We see also that such a rough estimation of the increase in resonance widths is compatible with the overall increase in the final cross section obtained  in the full calculation, which proves the validity of the interpretation.

\section{Concluding remarks}

We have investigated the role of the permanent electric dipole moment of  HCO$^+$ in the dissociative recombination of the ion with low energy electrons. We found that the inclusion of the dipole moment increases significantly the coupling between the $p\sigma$ and $p\pi^{\pm}$ states and only a little the coupling between $p\pi^+$ and $p\pi^-$.  This leads to an increase of the cross section for all collision energies below 1 eV by about 50$\%$. The obtained theoretical cross section agrees well with the experimental data of Ref. \cite{lepadellec97} for energies between $0.06$ and $0.3$ eV. Although, the cross section is also increased for low energies ($<0.06$ eV), the theoretical result is still below the experiment by about a factor of two. More theoretical and experimental work has to be done to clarify the reason for the remaining difference.

\begin{acknowledgments}
This work has been supported by the National Science Foundation under Grant
No. PHY-0427460 and Grant No. PHY-0427376, by an allocation of NERSC
supercomputing resources. VK also acknowledges the support from the RTRA network {\it Triangle de la Physique}. We would like to thank Maurice Raoult for useful comments and discussions.
\end{acknowledgments}

\bibliography{3bodyDR}

\end{document}